# The dependence of hydrophobic interactions on the solute size

Q. Sun[*]

*Key Laboratory of Orogenic Belts and Crustal Evolution, Ministry of Education, The School of Earth and Space Sciences, Peking University*

**ABSTRACT:**

Based on our recent study on physical origin of hydrophobic effects, this is applied to investigate the dependence of hydrophobic interactions on the solute size. As two same hydrophobic solutes are dissolved into water, the hydration free energy is determined, and the critical radius (Rc) is calculated to be 3.2 Å. With reference to the Rc, the dissolved behaviors can be divided into initial and hydrophobic solvation processes. These can be demonstrated by the molecular dynamics simulations on $C_{60}$-$C_{60}$ fullerenes in water, and $CH_4$-$CH_4$ molecules in water. In the association of $C_{60}$ fullerenes in water, with decreasing the separation between $C_{60}$ fullerenes, hydrophobic interactions can be divided into H1w and H2s hydrophobic processes, respectively. In addition, it can be derived that maximizing hydrogen bonding provides the driving force in the association of hydrophobic solutes in water.

[*] Corresponding author.

E-mail: QiangSun@pku.edu.cn

# 1. Introduction

In general, the hydrophobic effects are explained by the tendency of non-polar molecules or molecular surfaces to aggregate in an aqueous solution (or to be expelled from water into an aggregate). The hydrophobicity plays an important role in a variety of phenomena in aqueous environments such as micelle formation, protein folding and aggregation, lipid membrane formation, and molecular recognition. Therefore, many theoretical and experimental works have been devoted to investigate the mechanism of hydrophobic interactions.

In a landmark paper by Frank and Evans,[1] a first attempt at providing a detailed theory of the hydrophobic effect was made. These authors proposed that a hydrophobic solute in water promotes, in its surrounding, an enhanced hydrogen bond network of water, referred to as an "iceberg". Later, hydrophobicity was coined by Kauzmann [2] to describe the tendency toward adhesion between the non-polar groups in aqueous solution, which predicted that water near non-polar solutes was more "ordered" than bulk water. Additionally, another theoretical approach to hydrophobic hydration was based on the scaled particle theory (SPT),[3] which estimates the work necessary to create a spherical cavity in water. Using SPT, Stillinger [4] suggested that water solvates large non-polar molecules differently than small molecules. In recent years, the theoretical approach by Lum, Chandler, and Weeks (LCW) [5,6] provided a quantitative description of structural and thermodynamic aspects of hydrophobic hydration over the entire small to large length scale region. From the LCW theory of hydrophobic salvation,[5,6] it presents that the crossover between small and large regime occurs on nanometer length scale. In addition, a simple analytical model to account for the water hydrogen bonding in hydrophobic hydration was proposed by Southall and Dill [7] and Xu and Dill.[8] As expected, the model suggests the hydration of large non-polar solutes (much larger than a water molecule) to occur via a mechanism very different from that of smaller solutes.

The strength of hydrogen bonding is stronger than that of van der Waals interactions, water plays an important role in the process of hydrophobic effects. In our recent study,[9] based on the structural studies on water and air/water interface,[10-13] when a hydrophobic solute is dissolved into water, it mainly affects the structure of interfacial water (the topmost water layer at the solute/water interface). From this, the hydration free energy is derived, which is utilized to investigate the physical origin of hydrophobic effects. It can be found that hydrophobic effects are ascribed to the structural competition between hydrogen bondings in bulk water and those in interfacial water.[9]

In this work, based on our recent study on physical origin of hydrophobic effects,[9] this is applied to investigate the dependence of hydrophobic interactions on the solute size. As two hydrophobic solutes are dissolved into water, the hydration free energy is determined, the critical radius (Rc) is calculated to be 3.2 Å. Additionally, to investigate the mechanism of hydrophobic interactions, the molecular dynamics simulations are also performed on $C_{60}$-$C_{60}$ fullerenes in water, and $CH_4$-$CH_4$ molecules in water.

## 2 Molecular dynamics

### 2.1 Molecular dynamics simulations

In this study, the molecular dynamics simulations are conducted on $C_{60}$-$C_{60}$ fullerenes in water, and $CH_4$-$CH_4$ molecules in water using the program NAMD.[14] The simulations were performed in the isobaric-isothermal ensemble (NPT). The simulated temperature was kept at 300 K, employing moderately damped Langevin dynamics. The pressure was maintained at 1 atm using a Langevin piston.[15] The simulated box was 40Å×40Å×60Å for $C_{60}$-$C_{60}$ fullerenes in water, and 32Å×32Å×48Å for $CH_4$-$CH_4$ in water, respectively. Periodic boundary conditions were applied in the three directions of Cartesian space.

The empirical CHARMM force field [16] was used to describe inter-atomic interactions. The intermolecular three point potential (TIP3P) model [17] was employed to represent the water molecules. Non-bonded van der Waals interactions were smoothly switched to zero between 10 and 12 Å. The PME algorithm [18] was utilized to account for long-range electrostatic interactions. The equations of motion were integrated with a time step of 2 fs.

**2.2 Abf theory**

Hydrophobicity can generally be described in terms of the hydrophobic pair interactions that occur between solute particles, and can be calculated using a potential of mean force (PMF). In this work, PMF was calculated using NAMD with the adaptive biasing force (ABF) [19-24] extensions integrated in the Collective Variables module.[25] This method is a combination of probability density and constraint force methods, and is based on the thermodynamic integration of average force acting on coordinates.[19]

In the framework of ABF, the free energy along a transition coordinate can be seen as a potential resulting from the average force acting along the coordinate. In terms of generalized reaction coordinate $\xi$, the derivative of PMF can be expressed as,

$$\frac{dA(\xi)}{d\xi} = \left\langle \frac{\partial v(x)}{\partial \xi} - \frac{1}{\beta}\frac{\partial \ln|J|}{\partial \xi} \right\rangle_\xi = -\langle F_\xi \rangle_\xi \qquad (1)$$

where $|J|$ is the determinant of the Jacobian for the transformation from Cartesian to generalized coordinates, v(x) is the potential energy function, and $F_\xi$ is the instantaneous force. In the above equation, the first term of the ensemble average corresponds to the Cartesian forces exerted on the system, derived from the v(x) and the second term is a geometric correction arising from the change in the metric of the phase space due to the use of generalized coordinates.

In the ABF method, a biasing force opposing the actual force arising from system components is periodically applied to the reaction coordinate to generate what is effectively a random walk along the reaction coordinate (purely diffusive dynamics). The force applied along the reaction coordinate $\xi$, to overcome the PMF barriers is defined by,

$$F^{ABF} = \nabla_X \tilde{A}(\xi_0) = -\langle F_{\xi_0} \rangle_{\xi_0} \qquad (2)$$

where $\tilde{A}$ denotes the current estimate of the free energy and $\langle F_\xi \rangle_\xi$, the current average of $F_\xi$. As sampling of the phase space proceeds, the estimate $\nabla_x \tilde{A}$ is progressively refined.

In a simulation, the instantaneous force is calculated by dividing the reaction coordinate into discrete bins so that the average force exerted in the bin k is given by,

$$\overline{F_\xi}(N_{step}, k) = \frac{1}{N(N_{step}, k)} \sum_{i=1}^{N(N_{step}, k)} F_i(t_i^k) \qquad (3)$$

where N ($N_{step}$,k) is the number of samples collected in the bin k after $N_{step}$ simulation steps. $F_i(t_i^k)$ is the computed force at iteration i, and $t_i^k$ is the time at which the i*th* sample was collected in the bin k. For a sufficiently large $N_{step}$, $\overline{F}_\xi(N_{step},k)$ approaches the correct average force in each bin. Then, the free-energy difference, $\Delta A_\xi$, between the end point states can be estimated simply by way of summing the force estimates in individual bins,

$$\Delta A_\xi = -\sum_{i=1}^{M} \overline{F_\xi}(N_{step}, k) \delta\xi \qquad (4)$$

## 3 Discussions

**3.1 Hydrophobic interactions**

In principle, when a hydrophobic solute is dissolved into water, the thermodynamic functions may contain solute-solute, solute-solvent and solvent-solvent interaction energies, respectively.

$$\Delta G = \Delta G_{Water-water} + \Delta G_{Solute-water} + \Delta G_{Solute-solute} \tag{5}$$

The strength of hydrogen bonding in water is stronger than that of van der Waals interactions, therefore water should play an important role in the process of hydrophobic effects. To investigate the mechanism of hydrophobic interactions, it is necessary to study the structure of water, and the effects of the dissolved solute on water structure.

In our recent studies,[10-12] based on the dependence of OH vibrations on water molecular clusters, it can be found that, when three-dimensional hydrogen bondings appear, OH vibrations are mainly dependent on local hydrogen-bonded networks of a molecule, and hydrogen bondings beyond the first shell are weak. From this, different OH vibrations can be assigned to OH vibrations engaged in various local hydrogen bonding motifs. For ambient water, the main local hydrogen bonding motifs for a water molecule can be classified as DDAA (double donor-double acceptor), DDA (double donor-single acceptor), DAA (single donor-double acceptor), and DA (single donor-single acceptor).[10-12]

To date, many works have been devoted to investigate the structure of liquid water, and most models of water can be partitioned into two broad categories: (a) mixture models and (b) distorted hydrogen bond or continuum models.[26] According to Raman spectroscopic studies on ambient water,[12] a water molecule interacts with neighboring water molecules through various local hydrogen bonds, such as DDAA, DDA, DAA and DA. The hydrogen bonding of water is influenced by temperature, pressure, dissolved salt, and confined environments, which will be rearranged to oppose the changes of external conditions.

In fact, when a hydrophobic solute is dissolved into water, an interface appears between the solute and water. The OH vibrational frequency is mainly dependent on the local hydrogen bonding of a water molecule,[10-12] therefore the solute mainly affects the structure of topmost water layer at the interface (interfacial water). In comparison with bulk water, due to the truncations of hydrogen bondings, no DDAA (tetrahedral) hydrogen bonding can be found in the interfacial water,[13] which is closely related to the interfacial formation.

According to our recent work on the air/water interface,[13] after the ratio of interfacial water layer to volume is determined, this can be utilized to calculate the loss of DDAA hydrogen bonding, and applied to determine the Gibbs energy of interfacial water.

$$\Delta G_{Solute/water\,interface} = \Delta G_{DDAA} \cdot R_{Interfacial\,water/volume} \cdot n_{HB} \qquad (6)$$

where $R_{Interfacial\,water/volume}$ means the molecular number ratio of interfacial water layer to volume, and $n_{HB}$ is the average number of tetrahedral hydrogen bonding per molecule. For DDAA hydrogen bonding, $n_{HB}$ equals to 2. The $\Delta G_{DDAA}$ is the Gibbs free energy of DDAA (tetrahedral) hydrogen bonding.

In our previous work,[9] it is aimed at only a solute is dissolved into water. After the solute is regarded as an ideal sphere, the molecular number ratio of interfacial water layer to volume is calculated to be $4 \cdot r_{H2O}/\vec{R}$, where $r_{H2O}$ is the average radius of a H$_2$O molecule, and $\vec{R}$ is radius vector. Therefore, the hydration free energy can be determined as,

$$\Delta G_{Hydration} = \Delta G_{Water} + \Delta G_{Solute/water\,interface} = \Delta G_{Water} + \frac{8 \cdot \Delta G_{DDAA} \cdot r_{H2O}}{\vec{R}} \qquad (7)$$

where $\Delta G_{Water}$ is the Gibbs free energy of pure water.

In fact, this can reasonably be applied to investigate the hydrophobic interactions as two solutes are dissolved into water. From the above, when two solutes are embedded into water, the hydration free energy can be expressed as,

$$\Delta G_{Hydration} = \Delta G_{Water} + \Delta G_{Solute1/water\,interface} + \Delta G_{Solute2/water\,interface} = \Delta G_{Water} + \frac{8 \cdot \Delta G_{DDAA} \cdot r_{H2O}}{\vec{R}_{Solute-solute}} \qquad (8)$$

where $\vec{R}_{Solute-solute}$ is the separation between the two solutes.

In principle, the lower hydration free energy, the more thermodynamically stable. The hydration free energy is the sum of Gibbs energy of water and interfacial water. Therefore, the structural transition can be expected to take place when $\Delta G_{Water}$ equals to $\Delta G_{Solute/water\,interface}$,

$$\Delta G_{Water} = \Delta G_{Solute/water\,interface} \quad (R = R_c) \qquad (9)$$

where Rc is the critical radius. After the two same solutes are regarded as ideal spheres, the critical radius Rc can be determined to be 3.2 Å at 293 K and 0.1 MPa (Fig. 1).

According to our recent study on hydrophobic effects,[9] as two solutes are dissolved into water, the dissolved behaviors are dependent on the solute size. With reference to the Rc, the dissolved behaviors can be divided into initial and hydrophobic solvation processes, respectively. This means that hydrophobic effects can be expected only when the radius of solute is larger than 3.2 Å at ambient conditions.

In the process of hydrophobic solvation, the Gibbs free energy of bulk water is lower than that of interfacial water, $\Delta G_{Water} < \Delta G_{Solute/water\,interface}$ (both of them are negative). To maximize the strength of bulk water $|\Delta G_{Water}|$, this undoubtedly leads to the solutes tend to be accumulated in water in order to minimize $|\Delta G_{Solute/water\,interface}|$. Regarding to the mechanism of hydrophobic interactions, the driving force can be ascribed to originate from bulk water.

The hydrophobic effects can be explained by the tendency of non-polar molecules or molecular surfaces to aggregate in an aqueous solution. To investigate the mechanism of hydrophobic interactions, it is necessary to study the dependence of hydration free energy on the changes of surface area incurred by the association of the dissolved solutes. In the process of hydrophobic interactions, it is assumed that the solute is considered as a rigid, or there is no a change in the volume of the solute. In the association of the solute, this decreases the available surface area of solutes, which can be expressed as γ·Surface area, where γ is the ratio of surface area to volume of solutes. From this, the hydration free energy incurred by hydrophobic interactions can be expressed as,

$$\Delta G_{Hydration} = \Delta G_{Water} + \gamma \cdot \Delta G_{Solute/water\,interface} \qquad (10)$$

In principle, the lower hydration free energy (negative), the more thermodynamically stable. In the process of hydrophobic effects, the $\Delta G_{Water}$ is less than $\Delta G_{Solute/water\ interface}$ (negative). To be more thermodynamically stable, this may be fulfilled by minimizing $|\gamma \cdot \Delta G_{Solute/water\ interface}|$ in order to maximize the $|\Delta G_{Water}|$. This indicates that the strength of hydrophobic interactions may be orientation-dependent. In other words, when two solutes are aggregated, they prefer the specific direction to minimize the ratio of the surface area to volume of the solutes. In fact, this can be demonstrated by the molecular dynamics of graphite sheets in water,[27,28] CNTs in water.[28,29]

To investigate the thermodynamics in the process of hydrophobic interactions, the water-induced Gibbs free energy is expressed as,

$$\Delta G_{Water-induced} = -\Delta G_{Water} - \gamma \cdot \Delta G_{Solute/water\,interface} \qquad (11)$$

This can be used to investigate the thermodynamical changes incurred by the association of dissolved solutes. From the dependence of thermodynamics functions on the ratio of surface area to volume of solutes (Fig. 2), it can be found that, with decreasing the ratio (γ) of surface area to volume of solutes, this leads to the decrease of the Gibbs free energy (ΔG), the enthalpy (ΔH) and entropy (T·ΔS), respectively.

In our recent works,[9] as a hydrophobic solute is dissolved into water, it mainly affects the structure of interfacial water, which can be described as the effective radius of the solute. Of course, this is larger than the actual radius of the dissolved solute. In the process of hydrophobic interactions, due to the aggregation of the solutes, the water molecules between the solutes may be repelled into bulk water. Therefore, this may be accompanied with the changes of effective radius of dissolved solutes.

**3.2 Molecular dynamics simulations**

From ABF calculations, the PMF between $C_{60}$ fullerenes can be determined during the association of $C_{60}$ fullerenes in water (Fig. 3(a)). The first minimum in the PMF curve is referred to as the contact minimum. The second minima at separation of 13.0 Å is called the solvent-separated PMF, which corresponds to the separation at which one water molecular layer can enter the space between the two solutes (Fig. 4B). Additionally, another minimum at the separation of 16 Å can also be observed, which corresponds to double water molecular layers can be found between the $C_{60}$ fullerenes (Fig. 4A).

To investigate the water-induced contributions in the process of hydrophobic interactions, the PMFs between the solutes in vacuum are also calculated (Fig. 3). From these, the water-induced contributions to the PMF can be determined as,

$$\Delta G_{Wter-induced} = \Delta G_{Solute-solute\,in\,water} - \Delta G_{Solute-solute\,in\,vacuum} \qquad (12)$$

From figure 3, the water-induced PMF between $C_{60}$ fullerenes in water is completely different from that between $CH_4$ molecules in water. These are in agreement with other molecular simulations [29-36] on $C_{60}$ fullerenes in water, and $CH_4$ molecules in water. Of course, this indicates that the hydrophobic interactions are dependent on the size of solute.

From the PMF of the solutes in vacuum (Fig. 3(a)(b)), the radius of $C_{60}$ fullerene is 5.0 Å, and the radius of $CH_4$ is determined to be 1.9 Å. From the above discussion on hydrophobic interactions, as two same solutes are dissolved into water, in reference to Rc (3.2 Å), it can be divided into initial and hydrophobic solvation processes. The radius of $CH_4$ is less than Rc, the solutes may be dispersed in water, and there exists the repulsive force between them. Regarding to $C_{60}$-$C_{60}$ fullerenes in water, the radius of $C_{60}$ fullerene is larger than Rc, they will be aggregated in water.

Regarding to water-induced PMF between $C_{60}$ fullerenes in water, the strength of hydrophobic interactions is dependent to the separation between $C_{60}$ fullerenes. With decreasing the separation between $C_{60}$ fullerenes, this increases the water-induced PMF between $C_{60}$ fullerenes. From the above, the strength of hydrophobic interactions can reasonably be fitted as (Fig. 5),

$$\Delta G_{Water-induced} = a + b/(r-10) \qquad (13)$$

where r is the distance between the centers of $C_{60}$ fullerenes, r-10 is the separation between the $C_{60}$ fullerenes. Additionally, a and b are fitted parameters.

When the separation between $C_{60}$ fullerenes is larger than 11.2 Å, the water-induced PMF between $C_{60}$ fullerenes is reasonably described by the fitted equation (Fig. 5). However, as the separation between $C_{60}$ fullerenes being less than 11.2 Å, the water-induced PMF is obviously less than the strength of hydrophobic interactions of the fitted equation (Fig. 5). In combination with the dependence of hydration free energy on the ratio of surface area to volume (Fig. 2), it can be derived that this deviation between water-induce PMF and fitted equation should be closely related to the aggregation of the solutes.

When a solute is dissolved into water, it mainly affects the hydrogen bondings of interfacial water. In the process of hydrophobic effects, the solutes are attracted by each other. With decreasing the separation between the solutes, the water molecules between the solutes are repelled into bulk water. Therefore, water molecules can be found between the $C_{60}$ fullerenes at the separation of 16 Å (Fig. 4A) and 13 Å (Fig. 4B). This is different from the water molecular distribution as the separation is less than 11.2 Å (Fig. 4C), where no water molecules can be found at the axis between the centers of $C_{60}$ fullerenes. From this, $r_H$ can be derived in the association of solutes. This means that, as separation being less than $r_H$, no water molecules can be observed at the axis between the solutes. In reference to $r_H$, the hydrophobic interaction can be divided into H1w and H2s hydrophobic processes, respectively.

In fact, the structural difference between H1w and H2s hydrophobic processes can be reflected on the kinetics of $C_{60}$-$C_{60}$ fullerenes in water (Fig. 6). From the dependence of the separation between $C_{60}$ fullerenes on simulated time, in comparison with the slow decrease of separation between $C_{60}$ fullerenes, a sudden decrease can be found at the corresponding time of $r_H$. In addition, this difference can also be observed on the changes of molecular number changes of interfacial water and bulk water (Fig. 7). As two hydrophobic particles approach each other, a sudden drying transition or "evaporating" can be expected when the separation being less than a critical separation ($r_H$).

[1] H1w hydrophobic interaction (>$r_H$)

In this study, H1w means the water-related hydrophobic interactions. In the process of H1w hydrophobic interactions, water molecules can be found between the dissolved solutes. With decreasing the separation between the solutes, the water molecules between the dissolved solutes are repelled into bulk water. Therefore, in the association of dissolved solutes in water, the hydrophobic interactions are fulfilled by the rearrangement of water molecules between the solutes.

From the water-induced PMF between $C_{60}$ fullerenes in water, it can be derived that the strength of H1w hydrophobic interaction is inversely proportional to the separation between the $C_{60}$ fullerenes. From the above, this means that the strength of hydrophobic interactions should be orientation-dependent. In other words, when solutes are dissolved into water, the tendency of non-polar molecules or molecular surfaces to aggregate prefers specific direction to minimize the ratio of surface area of volume of solutes.

[2] H2s hydrophobic interactions ($<r_H$)

In fact, H2s hydrophobicity means the solute-related hydrophobic interactions. In the process of H2s hydrophobic interactions, no water molecules can be found at the axis between the dissolved solutes. In the association of dissolved solutes, this may be fulfilled by the molecular aggregation of pure dissolved solutes. Of course, this is different from H1w hydrophobic interactions discussed as above.

Regarding to the strength of H2s hydrophobic interactions, it is not only related to the separation between the solutes but also to the accumulation of surface area of solutes. As the separation between $C_{60}$ fullerenes is less than 11.2 Å, due to the aggregation of the solutes, this decreases the surface area of $C_{60}$ fullerenes available for interfacial water. Of course, this undoubtedly results in the decrease of the water-induced PMF between $C_{60}$ fullerenes.

Regarding to the water molecular number between $C_{60}$ fullerenes, a sudden decrease can be found in the H2s hydrophobic process (Fig. 7(b)). This is much like what would happen in a gas-liquid phase transition, or deweting transition. In fact, water detweting from nonpolar confinement has been observed in a number of simulation studies from nanotubes [37] to plates [38] and the interfaces between proteins [39] to collapsing polymers.[40]

For the sudden decrease on the separation and between $C_{60}$ fullerenes, this means that, in comparison with H1w process, an additional attractive force is imposed on the solutes in H2s process. In other words, this may be related to the changes of potential energy when the solutes approach each other. From the simulations, when the separation between $C_{60}$ fullerenes is less than 11.2 Å, it should be noted that the strength of van der Waals interactions between $C_{60}$ fullerenes in vacuum is stronger than the PMF of $C_{60}$ fullerenes in water. Therefore, with decreasing the separation between $C_{60}$ fullerenes, the solutes are undoubtedly affected by van der Waals interactions.

When the hydrophobic solutes are dissolved into water, the solutes are firstly attracted by the long-range hydrophobic interactions to be close to each other. In the H1w hydrophobic process (larger than $r_H$), the configurations presents a slow hydrophobic changes. However, when the separation being less than $r_H$ (H2s hydrophobic process), the dissolved solutes are affected by the short-range interactions, such as van der Waals interactions, and which leads to a sudden hydrophobic collapse (deweting transition).

From the ABF calculations on the PMF between $C_{60}$ fullerenes in water, the minima at separation of 16.0 Å and 13.0 Å correspond to two and one layer of hydrating water between the solutes, respectively. These can also be observed in the water-induced PMF of $C_{60}$ fullerenes in water, and the strength is inversely proportional to the separation between the solutes. Of course, this is completely different from hydrophobic interactions as the separation between $C_{60}$ fullerenes being less than 11.2 Å, at which no water can be found at the axis between $C_{60}$ fullerenes. In fact, this means that hydrophobic interactions may be closely related to the structural changes of the interfacial water (the topmost water layer) of the solutes. In other words, these indicate that the dissolved solute should mainly affect the hydrogen bonding of the interfacial water layer. This is in agreement with our recent studies on the structure of water.[11,12]

Hydrophobic effects can be explained by the tendency of non-polar molecules or molecular surfaces to aggregate in water. Therefore, in the association of dissolved solutes, this may be accompanied with the changes of the interfacial water layer and bulk water. From the molecular dynamics simulations on the $C_{60}$ fullerenes in water, with decreasing the separation between $C_{60}$ fullerenes, this decreases the ratio of interfacial water, but increases the ratio of bulk water, respectively (Fig. 7(a)(b)). These undoubtedly should be related to the structural difference between interfacial water and bulk water.

To investigate the origin of hydrophobic effects, it is necessary to study the changes of hydrogen bondings in the process of hydrophobic interactions. In this study, the geometrical definition of hydrogen bonding is utilized to determine the hydrogen bonds in water.[41] This definition makes use of both the oxygen-oxygen distance ($r_{OO}$) and the ∠OOH angle between two water molecules. If $r_{OO}$ and ∠OOH are less than 3.5 Å and 30°, a hydrogen bonding is considered to exist between two water molecules. From the molecular dynamics simulations, the hydrogen bonding number (average number of hydrogen bonds per water molecule, $n_{HB}$) can be calculated to investigate the structural changes of water.

In the association of $C_{60}$ fullerenes in water, it can be found that no obvious structural changes can be found in the structure of interfacial water and bulk water (Fig. 8). Due to the truncations of hydrogen bonding at the interface between solute and water, the hydrogen bonding number of interfacial water is less than that of bulk water (Fig. 8). Therefore, the structure of bulk water should be more thermodynamically stable than that of interfacial water. This can be applied to understand that maximizing bulk water may provide the driving force in the association of $C_{60}$ fullerenes in water.

From the changes of hydrogen bonding in the association of $C_{60}$ fullerenes in water, the dissolved solute mainly affect the structure of water molecules in the first shell, and this decreases the hydrogen bonding number in water. The structural origin of hydrophobic effects is due to the strength of bulk water is stronger than that of interfacial water. In other words, the hydrophobicity originates from the structural competition between interfacial water and bulk water, and maximizing hydrogen bonding provides the driving force in the association of hydrophobic solutes in water (Fig. 8). Additionally, this is in agreement with our recent study [9] on the physical origin of hydrophobic effects.

**4 Conclusions**

Based on our recent work on physical origin of hydrophobic effects, this is applied to investigate the dependence of hydrophobic interactions on the solute size. Additionally, to investigate the mechanism of hydrophobic interactions, the molecular dynamics simulations are performed on $C_{60}$-$C_{60}$ fullerenes in water, and $CH_4$-$CH_4$ molecules in water. From this study, the following conclusions can be derived,

(1) As two hydrophobic solutes are dissolved into water, after the hydration free energy is determined, the Rc (critical radius) is calculated to be 3.2 Å at 293 K and 0.1 MPa. In reference to the Rc, the dissolved behaviors are dependent on the solute size, and can be divided into initial and hydrophobic solvation processes, respectively.

(2) From the ABF calculations on $C_{60}$-$C_{60}$ fullerenes in water, the water-induced PMF can be determined. With decreasing the separation between $C_{60}$ fullerenes in water, in reference to $r_H$, it can be divided into H1w and H2s hydrophobic interactions, respectively.

(3) When a solute is embedded into water, it mainly affects the structure of interfacial water of the solute. Regarding to the mechanism of hydrophobic interactions, it is due to the strength of bulk water is stronger than that of interfacial water. Therefore, the hydrophobic interactions can be ascribed to the structural competition between interfacial water and bulk water, and maximizing hydrogen bonding provides the driving force in the association of hydrophobic solutes in water.

**Acknowledgements**

This work is supported by the National Natural Science Foundation of China (Grant Nos. 41373057).

**Fig. 1.** The dependence of hydration free energy on the solute size. As two same solutes are dissolved into water, the Rc is 3.2 Å at 293 K and 0.1 MPa.

**Fig. 2.** The dependence of hydration free energy on the ratio ($\gamma$) of surface area to volume of solutes. In the association of solutes, the decrease of the ratio $\gamma$ from 0.8 to 0.4 may decrease the Gibbs free energy ($\Delta G$), the enthalpy ($\Delta H$) and the entropy ($T \cdot \Delta S$).

**Fig. 3.** (a) The PMF between $C_{60}$ fullerenes in water and vacuum. (b) The PMF between $CH_4$ dimer in water and vacuum. (c) The water-induced PMF between $C_{60}$ fullerenes in water. (d) The water-induced PMF between $CH_4$ dimer in water.

**Fig. 4.** Snapshots of water molecules between $C_{60}$ fullerenes with the different separation between $C_{60}$ fullerene of 16 Å (A), 13 Å (B), and 11.2 Å (C). The interfacial water layer of $C_{60}$ fullerene (D).

**Fig. 5.** The water-induced PMF between $C_{60}$ fullerenes in water at 300 K and 1 bar. In reference to $r_H$ (dashed line), the hydrophobic interactions are divided into H1w and H2s hydrophobic processes. Regarding to the strength of H1w hydrophobic interaction, it is fitted as,

ΔG=-3.35+3.80/(r-10). In the H2s hydrophobic process, the strength of hydrophobic interactions at various ratio (0.8, 0.6, 0.4) of surface area to volume are drawn in squares.

**Fig. 6.** The kinetic changes of the separation between $C_{60}$ fullerenes at 300 K and 1 bar.

**Fig. 7.** The changes of interfacial water (a) and bulk water (b) in the association of $C_{60}$ fullerenes in water at 300 K and 1 bar. The corresponding time of $r_H$ is drawn in dashed line.

**Fig. 8.** The hydrogen bonding number per water molecule in interfacial water, bulk water (a) and total water (b) at 300 K and 1 bar. The fitted line is also shown.

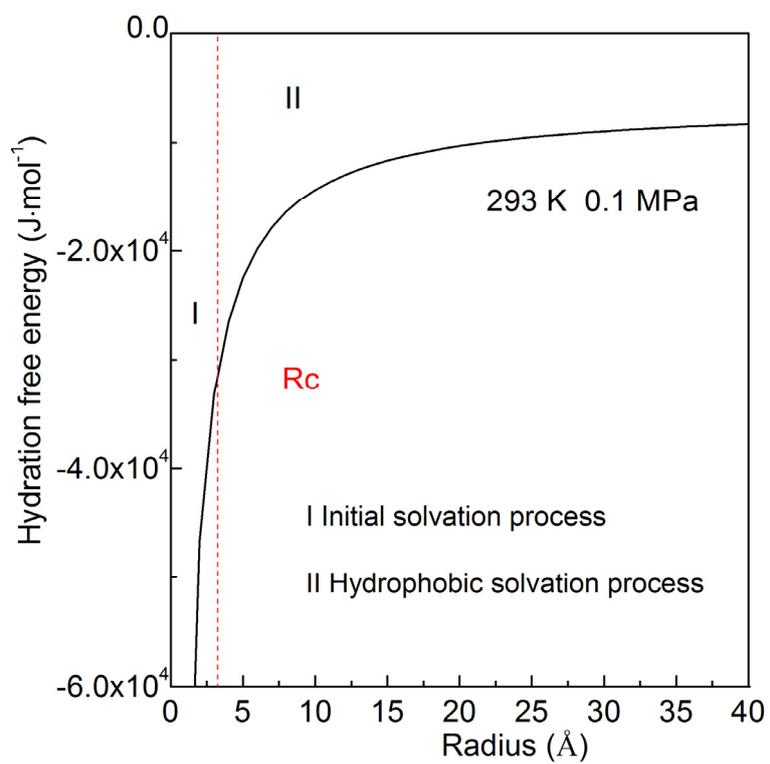

Fig. 1.

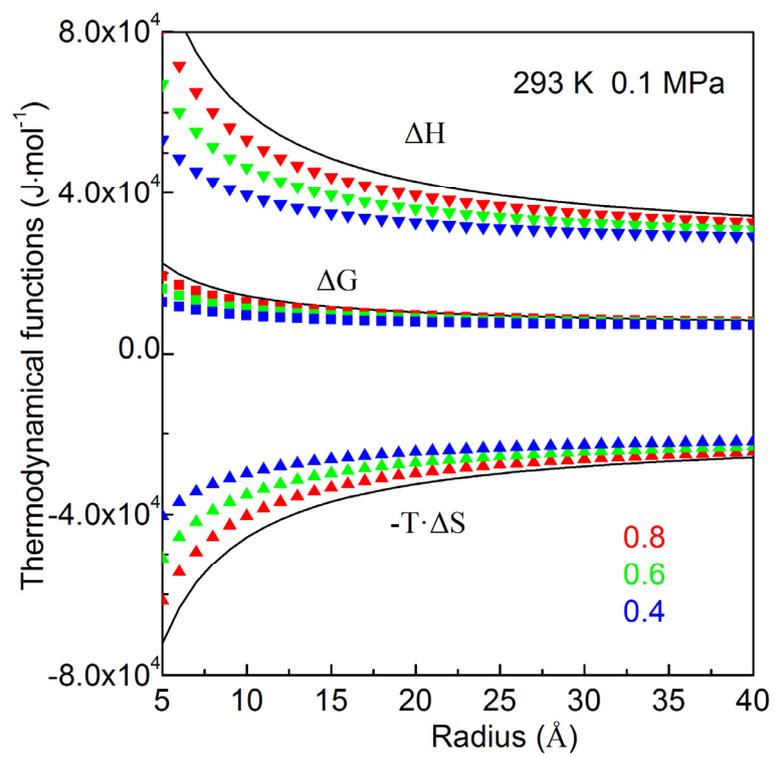

Fig. 2.

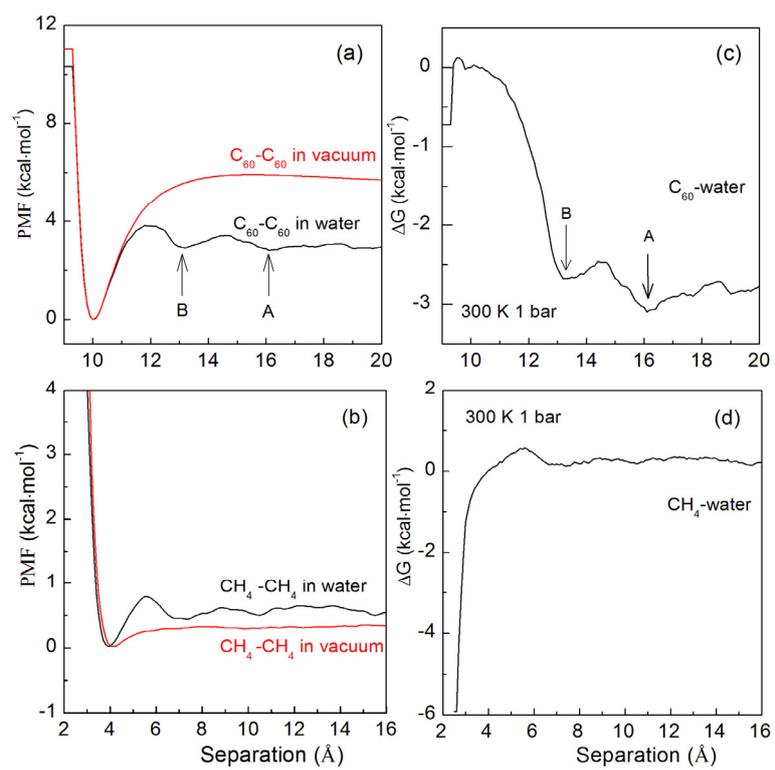

Fig. 3.

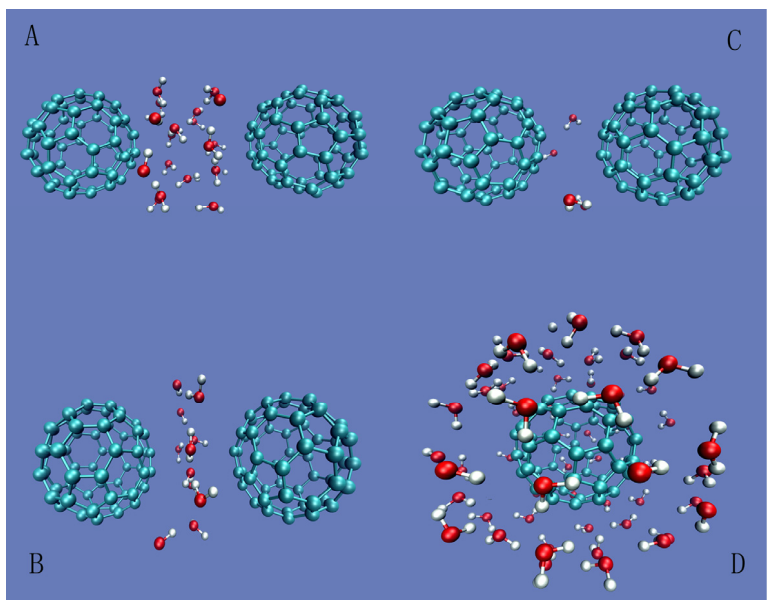

Fig. 4.

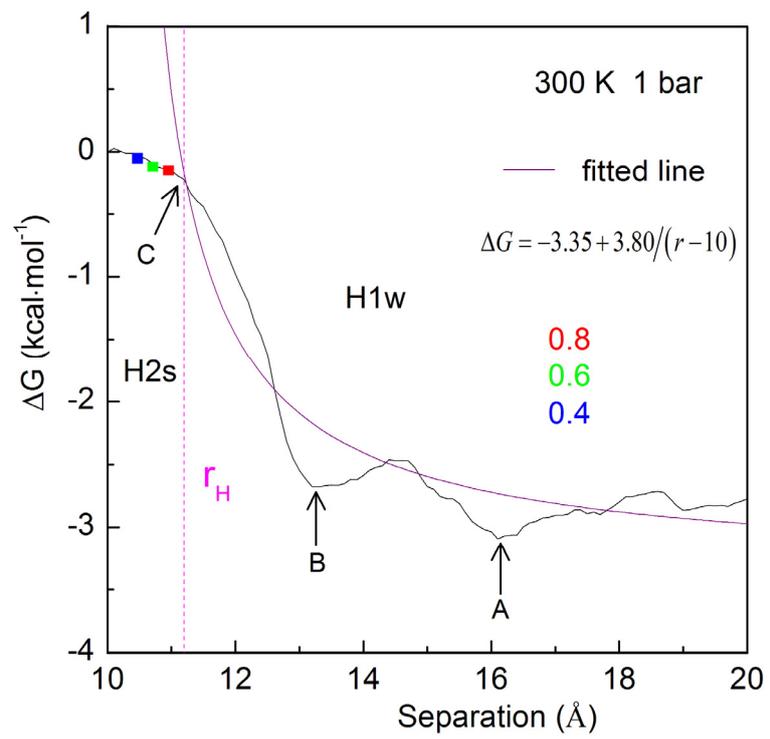

Fig. 5.

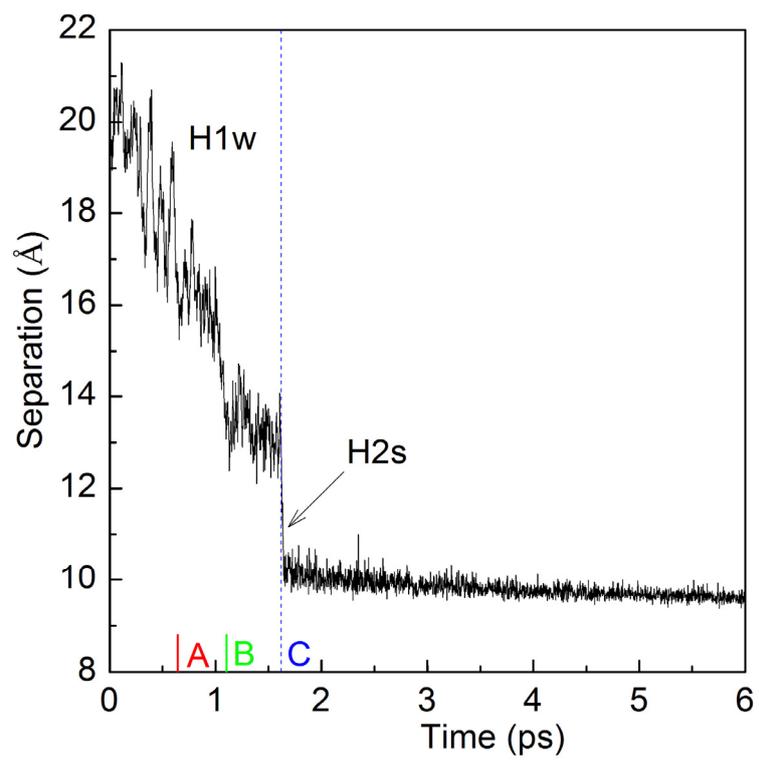

Fig. 6.

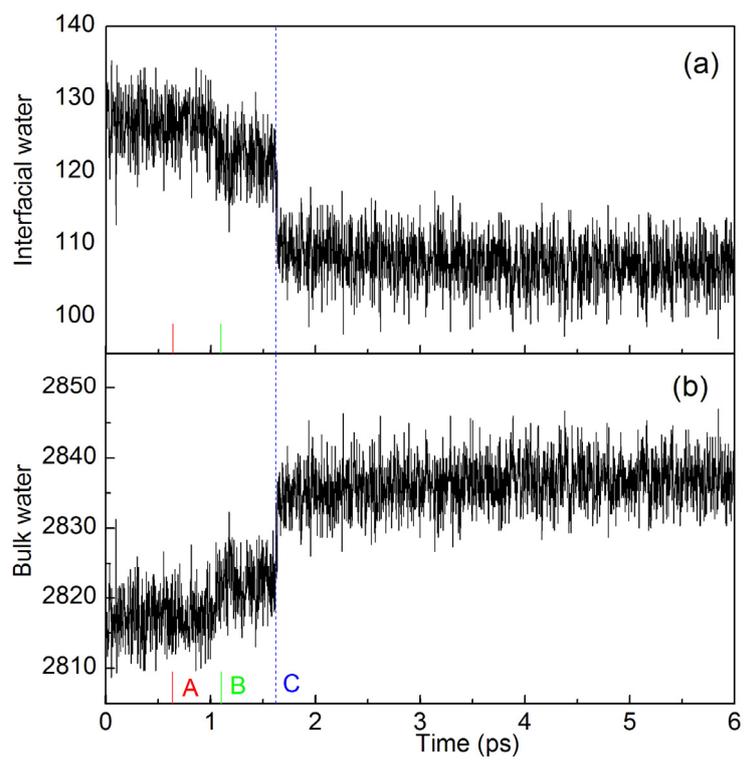

Fig. 7.

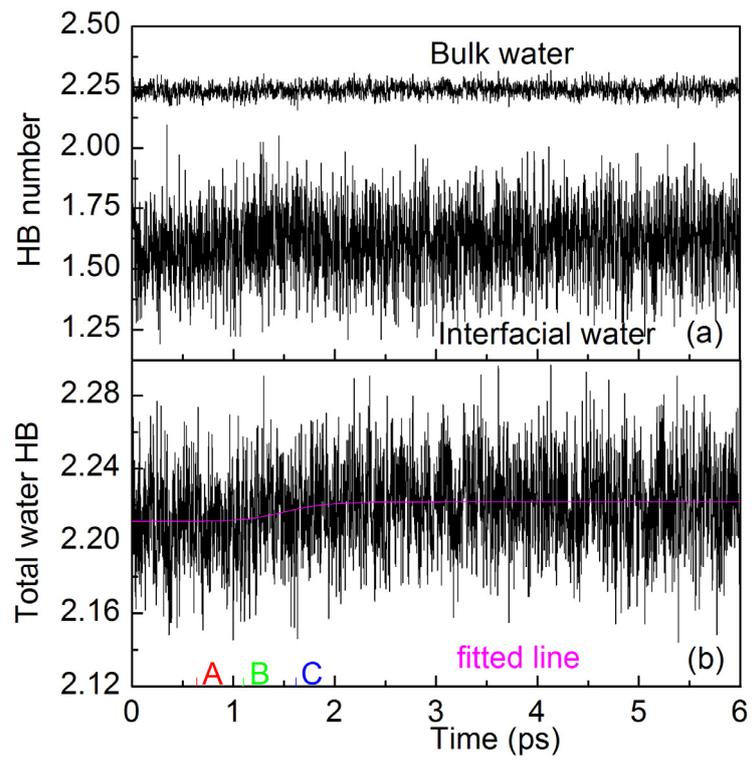

Fig. 8.